%% file: main.tex
\RequirePackage{lineno}
\documentclass[aps,prl,twocolumn,groupedaddress,amsmath,amssymb]{revtex4-2}

\usepackage{amssymb}
\usepackage{overpic}
\usepackage{graphicx}
\usepackage{subfigure}
\usepackage{dcolumn}% Align table columns on decimal point
\usepackage{bm}% bold math
\usepackage{epsfig}
\usepackage{xcolor}
\usepackage{enumerate}
\usepackage{booktabs}
\usepackage{multirow}
\usepackage{braket}
\usepackage{tikz}
\usepackage{comment}

\usepackage[colorlinks,
            urlcolor=blue,
            linkcolor=blue,
            anchorcolor=blue,
            citecolor=blue]{hyperref}
\newcommand{\BESIIIorcid}[1]{\href{https://orcid.org/#1}{\hspace*{0.1em}\raisebox{-0.45ex}{\includegraphics[width=1em]{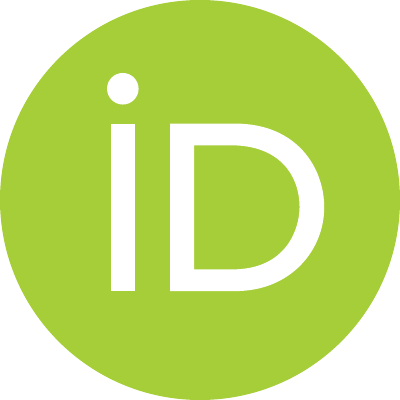}}}} 

\begin{document}

\newcommand{\jpsi}{J/\psi}
\newcommand{\pip}{\pi^+}
\newcommand{\pin}{\pi^-}
\newcommand{\pio}{\pi^0}
\newcommand{\g}{\gamma}
\newcommand{\gev}{GeV/c$^2$}
\newcommand{\mev}{MeV/c$^2$}
\newcommand{\ar}{\rightarrow}
\newcommand{\ks}{K_S^{0}}
\newcommand{\etap}{\eta^\prime}
\let\oldequation\equation
\let\oldendequation\endequation
\renewenvironment{equation}{\linenomathNonumbers\oldequation}{\oldendequation\endlinenomath}

%Title of paper
\title{\Large \boldmath \bf Observation of a threshold enhancement in the $\pi^+\pi^-$ spectrum in $\psi(3686) \rightarrow \pi^{+}\pi^{-}J/\psi$ decays}

\author{
\begin{small}
 \begin{center}
\input{authorlist_2025-07-02}
\end{center}
\end{small}
}

% \linenumbers

%####################################     abstract
\begin{abstract}
Based on the $(2712.4\pm14.4)\times 10^{6}$ $\psi(3686)$ events collected with the BESIII detector, we present a high-precision study of the $\pi^+\pi^-$ mass spectrum in $\psi(3686)\rightarrow\pi^{+}\pi^{-}J/\psi$ decays. A clear resonance-like structure is observed near the $\pi^+\pi^-$ mass threshold for the first time. A fit with a Breit-Wigner function yields  a mass of $285.6\pm 2.5~{\rm MeV}/c^2$  and a width of $16.3\pm 0.9~{\rm MeV}$ with a statistical significance exceeding 10$\sigma$. 
To interpret the data, we incorporate final-state interactions (FSI) within two theoretical frameworks: chiral perturbation theory (ChPT) and QCD multipole expansion (QCDME).  ChPT describes the spectrum above  0.3 GeV/$c^2$ but fails to reproduce the threshold enhancement. In contrast, the QCDME model—assuming the  $\psi(3686)$ is an admixture of S- and D-wave charmonium—reproduces the data well. The pronounced dip near 0.3 GeV/$c^2$  offers new insight into the interplay between chiral dynamics and low-energy QCD.

\end{abstract}

\maketitle

Charmonium, the bound state of a charm quark and its antiquark, is the QCD analogue of positronium. Unlike positronium, whose properties are governed primarily by quantum electrodynamics, charmonium is shaped by the strong interaction and therefore provides one of the simplest laboratories for probing QCD. Progress in understanding quantum chromodynamics has been driven to a large extent by detailed studies of charmonium and its heavier counterpart, bottomonium.

Di-pion transitions in heavy quarkonium decays play an essential role in testing theoretical models for a precise description of the $\pi\pi$ spectrum, thereby elucidating the internal structure of these states and the coupling of light quarks to heavy degrees of freedom~\cite{Voloshin:2006ce}.

The charmonium states $\psi(3686)$ and $\psi(3770)$ are well established and are conventionally assigned as the $2^3 S_1$ and $1^3D_1$ states, respectively.
However, the predicted leptonic width of $\psi(3770)$ is smaller than the experimental measurement by an order of magnitude~\cite{Kuang:2006me}, which inspires  the hypothesis that both states are mixtures of  $2^3 S_1$  and $1^3D_1$ components.
Transitions between charmonium states thus probe heavy-quark dynamics at short distances and simultaneously test the nature of these states. In particular, the most dominant decay of $\psi(3686)$, $\psi(3686)\rightarrow\pi\pi J/\psi$, is assumed to proceed via a two-step process: the emission of two gluons followed by their hadronization into a pair of pions. Owing to the small mass difference between the initial and final charmonia, the emitted gluons are soft and lie outside the perturbative regime, prompting intense theoretical interest in modeling the $\pi\pi$ mass spectrum~\cite{Kuang:2006me}.  

Experimentally, $\psi(3686)\rightarrow\pi\pi J/\psi$ has been studied extensively in $e^+e^-$  annihilation using a variety of theoretical models~\cite{Abrams:1975zp,BES:1999guu,CLEO:2008kwj,e760,besiii}. High-precision measurements of the di-pion system, however, remain scarce despite their importance for understanding $\psi(3686)-\psi(3770)$ mixing and the underlying decay dynamics. 

Moreover, in the soft-pion limit, where the momentum of one pion approaches zero, the amplitude of $\psi(3686)\to\pi\pi J/\psi$~\cite{Pham:1975zq} is expected to vanish, as required by the Adler zero condition~\cite{Adler:1964um}. This fundamental prediction, which has yet to be verified experimentally~\cite{Abrams:1975zp,BES:1999guu,CLEO:2008kwj,e760,besiii}, would provide direct evidence for the Nambu–Goldstone nature of the pion and offer crucial insight into chiral symmetry breaking.
Decays into two pions are also the leading channels for a Higgs-mixed scalar and hadrophilic scalar~\cite{Bezrukov:2009yw, Batell:2018fqo, Gan:2020aco}.

%In this Letter, we report the first observation of a distinct resonance-like structure
In this Letter, we report the first observation of a distinct threshold enhancement near the $\pi\pi$ threshold using $3.7\times 10^7$ $\psi(3686)\rightarrow\pi\pi J/\psi$ events selected from $(2712.4\pm14.4)\times 10^{6}$ $\psi(3686)$ events collected with the  BESIII detector~\cite{BESIII:2024lks}. A quantitative precision analysis is performed to characterize the structure and elucidate its decay dynamics.

The BESIII detector records symmetric $e^+e^-$ collisions provided by the BEPCII storage ring~\cite{Yu:2016cof} in the center-of-mass energy range from 1.84 to 4.95~GeV, which is described in detail in Refs.~\cite{BESIII:2009fln, Jiao:2020dqs, Zhang:2022bdc, Cao:2020ibk}. 

Signal candidates for the process $\psi(3686) \rightarrow \pi^+\pi^- J/\psi$, with $J/\psi \rightarrow \ell^+\ell^-$ ($\ell = e$ or $\mu$), are selected by requiring exactly four good charged tracks in an event with zero net charge. All charged tracks must be reconstructed within the acceptance of the Multilayer Drift Chamber (MDC).
From the four tracks, the two with the lower momenta in the laboratory frame (each required to be less than 1 GeV/$c$) are assigned as the $\pi^+$ and $\pi^-$ candidates. The remaining two tracks, which required to be larger than 1 GeV/$c$, are treated as the lepton ($e^\pm$ or $\mu^\pm$) candidates from the $J/\psi$ decay. No requirement is made on any photons that may be reconstructed.
To suppress background from direct $e^+e^-$ annihilation, the polar angle $\theta$ (with respect to the MDC symmetry axis) is required to satisfy $|\cos\theta| < 0.8$ for lepton candidates and $|\cos\theta| < 0.93$ for pion candidates. For each track, the distance of closest approach to the interaction point must be less than 10 cm along the beam ($z$) axis and less than 1 cm in the transverse plane.
Lepton identification is performed using the ratio of the energy deposit in the electromagnetic calorimeter to the track momentum ($E/p$). A track is identified as a muon candidate if $E/p < 0.26$, and as an electron candidate if $E/p > 0.8$. To further suppress background from photon conversions, the momentum of each lepton candidate is required to satisfy $|P_\ell - 1.56| < 0.17$ GeV/$c$, where 1.56 GeV/$c$ is the expected momentum for a lepton from $J/\psi \rightarrow \ell^+\ell^-$ in $J/\psi$ center-mass-system.
A five-constraint (5C) kinematic fit is performed on each candidate event, imposing overall energy-momentum conservation and constraining the invariant mass of the lepton pair to the nominal $J/\psi$ mass. Candidate events are retained only if the fit converges with $\chi^2_{\rm 5C} < 100$.

\begin{figure}[htbp]
    \centering
    \includegraphics[width=0.5\textwidth]{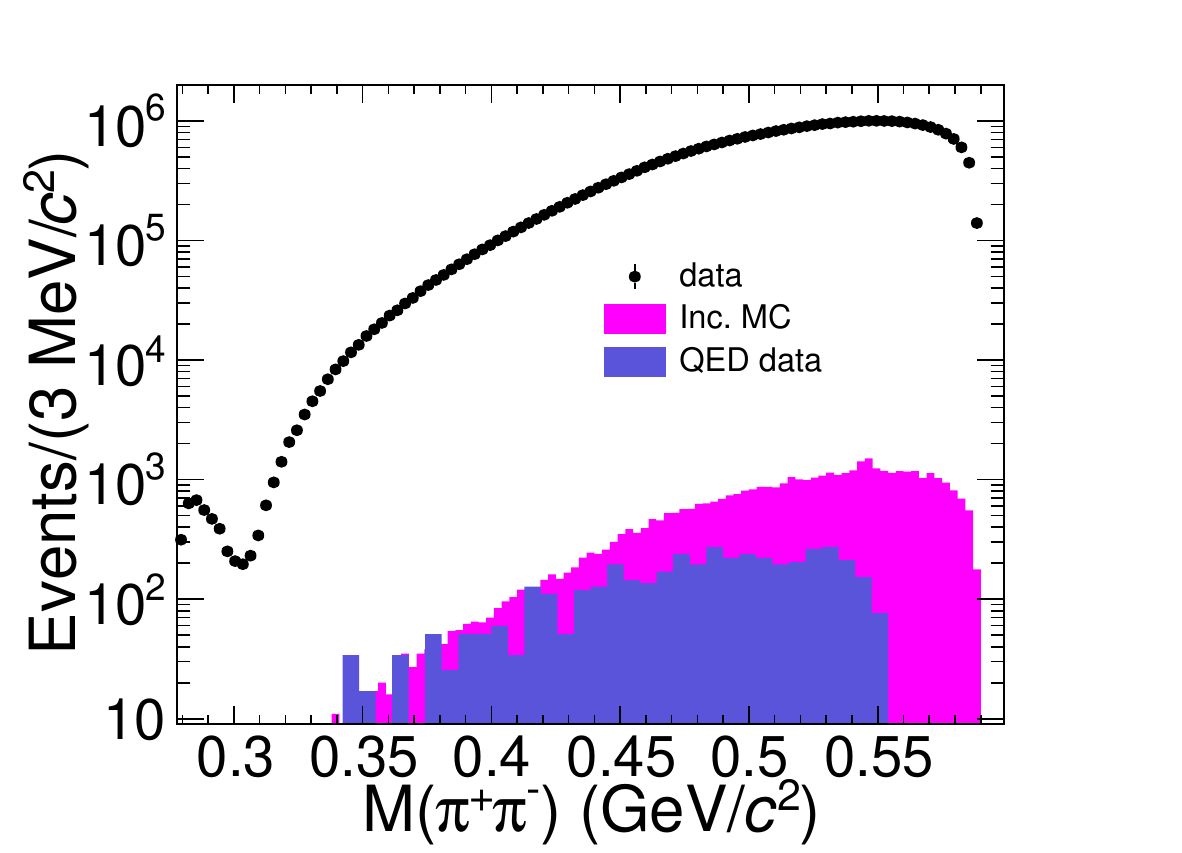}
    \caption{
    ${\rm M}(\pi^+\pi^-)$ distribution for $\psi(3686)\rightarrow\pi^+\pi^-J/\psi$ events. The dots with error bars are data. The violet-shaded histogram is the inclusive decay MC. The blue-shaded histogram is the data sample at $\sqrt{s}=3.65~{\rm GeV}$. 
    }
    \label{fig:bkgpipi}
\end{figure}

After applying the aforementioned selection criteria, a total of \(3.7 \times 10^7\) \(\psi(3686) \rightarrow \pi^{+}\pi^{-}J/\psi\) candidates are selected from both the \(J/\psi \rightarrow e^+e^-\) and \(J/\psi \rightarrow \mu^+\mu^-\) decay channels. The corresponding \(\pi^+\pi^-\) invariant mass spectrum is presented in Fig.~\ref{fig:bkgpipi}, which shows a pronounced enhancement near the \(\pi^+\pi^-\) mass threshold.

To ensure this threshold enhancement does not originate from a background contribution, extensive background studies are performed using both Monte Carlo (MC) simulations and data, which includes both the production of the $\psi(3686)$ resonance generated with {\sc lundcharm}~\cite{Chen:2000tv, Yang:2014vra} and {\sc evtgen}~\cite{Ping:2008zz}, the ISR production of the $J/\psi$ and the continuum processes incorporated in {\sc kkmc}~\cite{Jadach:2000ir}. Possible background contributions are investigated with an inclusive MC sample of 2.7 billion \(\psi(3686)\) decays. The primary background source is identified as \(\psi(3686) \rightarrow \pi^+\pi^- J/\psi\), \(J/\psi \rightarrow \pi^+\pi^-\). Additionally, the background from non-resonant \(e^+e^-\) annihilation is estimated using a data sample collected at \(\sqrt{s} = 3.65~\text{GeV}\) with an integrated luminosity of 445.5 pb\(^{-1}\). The overall background contamination rate is estimated to be approximately 0.02\%, and none of the background sources produce a peaking structure near the \(\pi^+\pi^-\) mass threshold, as illustrated in Fig.~\ref{fig:bkgpipi}.

\begin{figure}[htbp]
    \centering
    \includegraphics[width=0.5\textwidth]{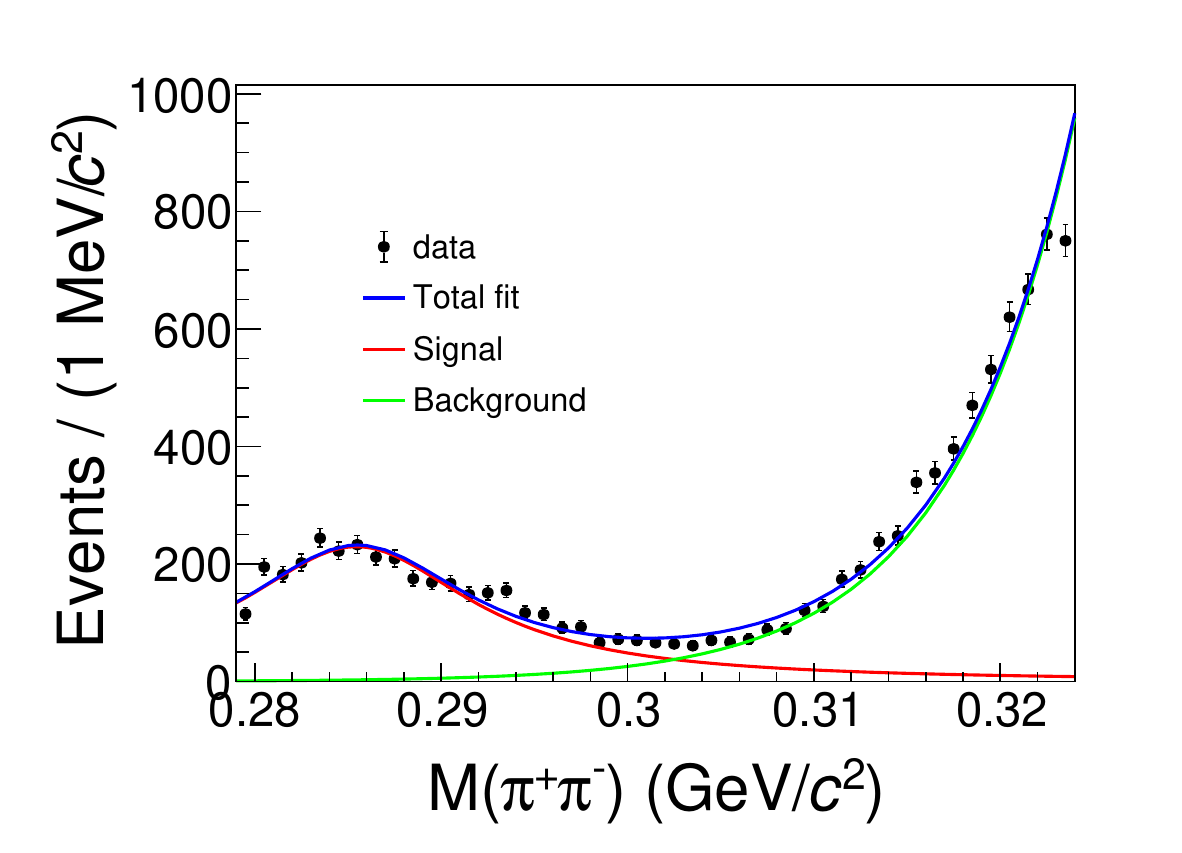}
    \caption{The fits to the ${\rm M}(\pi^+\pi^-)$ distribution. The dots with error bars represent data, the blue line represents the total fit, the red dotted line represents the signal, and the green line represents the non peaking background.
    }
    \label{fig:pipifit}
\end{figure}

An unbinned maximum likelihood fit is performed to the $\pi^+\pi^-$ invariant mass spectrum in the range of $[0.279, 0.325]~({\rm GeV}/c^2)$. The signal from the resonant hypothesis $\psi(3686) \rightarrow X J/\psi$ ($X \rightarrow \pi^+\pi^-$) is modeled by a Breit–Wigner (BW) function, convolved with a Gaussian function to describe the resolution, and multiplied by the detection efficiency. The contribution from the left tail of the broad structure above 0.3 GeV/$c^2$ is modeled by an exponential function.
As shown in Fig.~\ref{fig:pipifit}, the fit with $\chi^2$/ndf = 1.5 yields $3758.6 \pm 72.1$ signal events, with a mass of $(285.5 \pm 2.6)~\text{MeV}/c^2$ and a width of $(16.4 \pm 0.8)~\text{MeV}$, where the uncertainties are statistical only. Although narrow, this structure has a width corresponding to a lifetime orders of magnitude shorter than that of pionium, the $\pi^+\pi^-$ Coulomb bound state observed by DIRAC~\cite{DIRAC:2005hsg}. Pionium decays predominantly into $\pi^0\pi^0$ with a measured lifetime on the order of $10^{-15}$ seconds, whereas the observed width here implies a much shorter lifetime, ruling out a conventional pionium interpretation.

Detector resolution and efficiency distort the measured spectrum. To confront theoretical models with data, we unfold the binned spectrum (background subtracted) using a response matrix obtained from signal MC. The iterative Bayesian method~\cite{DAgostini:1994fjx, Richardson:1972hli, Lucy:1974yx} implemented in {\sc RooUnfold} software package~\cite{Brenner:2019lmf, Adye:2011gm} is employed with four iterations.

We perform a two-dimensional binned maximum-likelihood fit to the unfolded spectrum, using the $\pi^+\pi^-$ invariant mass in the range of $[0.279, 0.582]~{\rm GeV}/c^2$ and the $\pi^+$ helicity angle, $\cos\theta_{\pi^+}$ within $[-1,1]$, as the fit variables. Here, $\cos\theta_{\pi^+}$ is defined as the angle between the momentum of the $\pi^{+}$ in the rest frame of the $\pi^{+}\pi^{-}$ system and the momentum of the $\pi^{+}\pi^{-}$ system in the rest frame of the initial $\psi(3686)$.
The fits are conducted within two theoretical frameworks: chiral perturbation theory (ChPT)~\cite{Mannel:1995jt, Guo:2004dt, Chen:2015jgl} and QCD multipole expansion (QCDME)~\cite{Kuang:2006me,Kuang:2012wp}. Data from both the $J/\psi\rightarrow e^{+}e^{-}$ and $J/\psi\rightarrow \mu^{+}\mu^{-}$ decay channels are included and fitted simultaneously.

Within the framework of ChPT~\cite{Mannel:1995jt, Guo:2004dt, Chen:2015jgl},  the differential decay rate is
\begin{equation}
    \frac{d\Gamma}{dm_{\pi\pi}~d\cos{\theta_{\pi^+}}}=\frac{m_{\pi\pi}\sigma_{\pi}|\bf{q}|}
    {64\pi^{3}m_{\psi}^{2}}|\mathcal{M}_S+\mathcal{M}_D|^{2},
\end{equation}
where $|{\bf q}|=\frac{1}{2m_{\psi}}\lambda^{1/2}(m_{\psi}^{2}, m_{J/\psi}^{2}, m^{2}_{\pi\pi})$ and $\sigma_{\pi}=\sqrt{1-\frac{4m_{\pi}^{2}}{m_{\pi\pi}^2}}$. Here, $\mathcal{M}_S$ and $\mathcal{M}_D$ are the S-wave and D-wave amplitudes, respectively. The S- and D-wave amplitudes are $\mathcal{M}_S=[\frac{g_{0}}{2}(m_{\pi\pi}^2-2m_{\pi}^{2})+g_{1}E_{\pi^{+}}E_{\pi^{-}} ]\epsilon_{J/\psi}\cdot\epsilon^{*}_{\psi'} $, $\mathcal{M}_D=g_{2}[p_{\pi^{+}\mu}p_{\pi^{-}\nu} + p_{\pi^{+}\nu}p_{\pi^{-}\mu} ]\epsilon_{J/\psi}^{*\mu}\cdot\epsilon^{\nu}_{\psi'}$, 
where $E_{\pi^{(\pm)}}$ is the energy of pion, $p_{\pi^{(\pm)}}$ is the four momentum of pion, $g_{i}$ denotes the phenomenological constants, while $\epsilon_{J/\psi}$ and $\epsilon_{\psi'}$ are the polarization vector of $J/\psi$ and $\psi(3686)$, respectively.

\begin{figure}[htbp]
    \centering
    \subfigure{
        \begin{overpic}[page=1, width=0.5\textwidth]{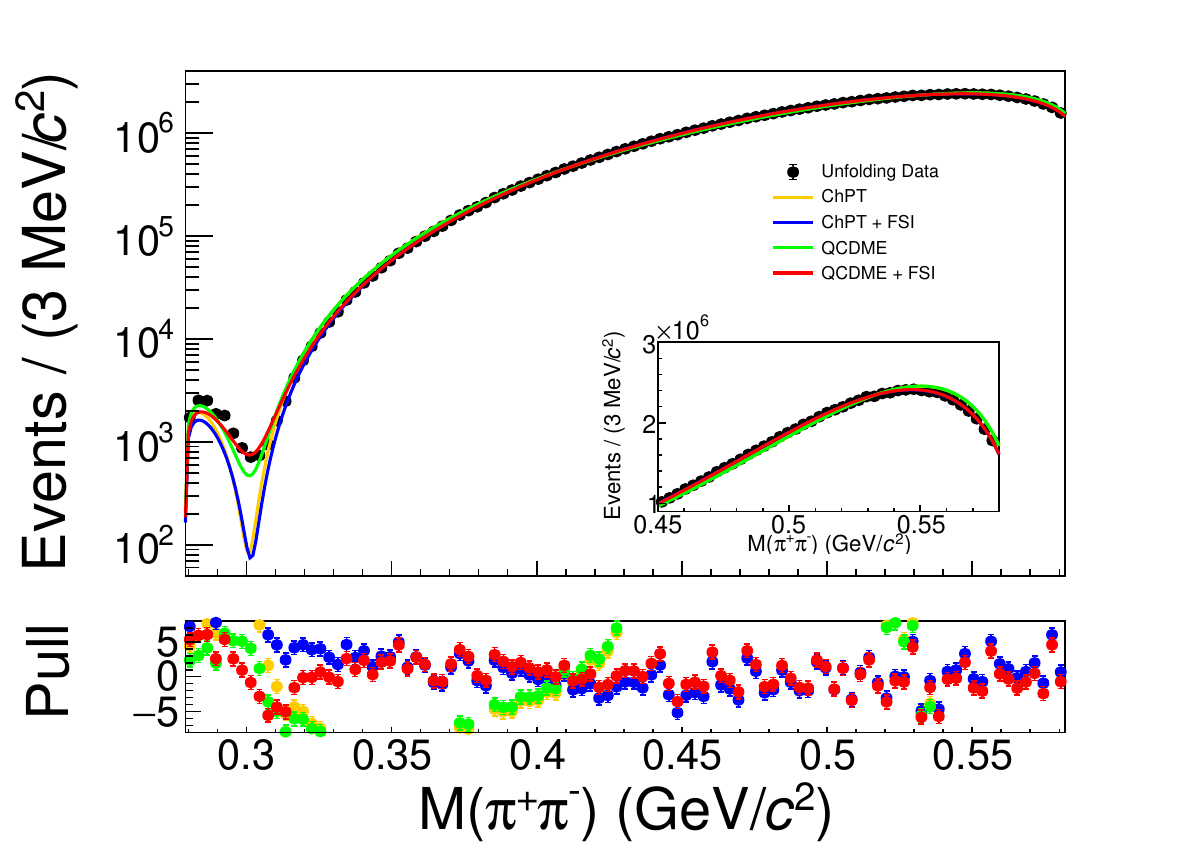}
        \put(18,55){\bf (a)}
        \end{overpic}
    }
    \subfigure{
        \begin{overpic}[page=2, width=0.5\textwidth]{compare.pdf}
        \put(18,55){\bf (b)}
        \end{overpic}
    }
    \caption{The fitting results of (a) $M(\pi^+\pi^-)$ and (b) $\cos{\theta_{\pi^+}}$ with four theories. The dots with error bars are for the unfolded data, which are corrected  for both efficiency and resolution effect, the yellow curve is the ChPT theory fitting results, the green curve is the ChPT theory with FSI fitting results, the blue curve is the QCDME fitting result and the red curve is the QCDME with FSI fitting result. The pull distribution is derived based on the QCDME with FSI fitting result and the data.}
    \label{fig:totalfit}
\end{figure}

A simultaneous fit to the $\pi^+\pi^-$ mass spectrum and the $\cos\theta$ angular distribution, varying the ratios $g_{1}/g_{0}$ and $g_2/g_0$, yields significant discrepancies in both the low- and high-mass regions (Fig.~\ref{fig:totalfit}).
Incorporating final-state interactions within the chiral unitary approach (CHUA)~\cite{Guo:2004dt, Liu:2012dv, Oller:2000ma}, the leading-order S-wave amplitude is corrected to be
\begin{equation}
\label{eq2}
    \mathcal{M}_S'=\mathcal{M}_S\cdot(1+  \frac{1}{16\pi^{2}}[ \gamma+\sigma_{\pi}\ln{ \frac{\sigma_{\pi}-1}{\sigma_{\pi}+1} } ] \cdot t_{\pi\pi})
\end{equation}
where $t_{\pi\pi} = (m_{\pi\pi}^2-m_\pi^2)/F_{\pi}$ and $F_\pi$ is the pion decay constant. Here $\gamma$ is a subtraction constant that regularizes the loop. 

An alternative fit is then performed, yielding, 
$ g_{1}/g_0 = -0.087 \pm 0.0006$,
$ g_{2}/g_0 = (0.01 \pm 1.20)\times 10^{-7}$, 
$\gamma = -3.711 \pm 0.002$. 
Based on the fitted projection of the $\pi^+\pi^-$ mass spectrum shown in Fig.~\ref{fig:totalfit}, the description of the data in the high-mass region around 0.55~GeV/$c^2$ is significantly improved, with the $\chi^2$/ndf decreasing from 310.5 without FSI to 16.1 when FSI is included.

In accordance with QCDME, the $\psi(3686)$ is modeled as an admixture of $\psi(2S)$ and $\psi(1D)$ states. The differential decay width can be described, following Refs.~\cite{Kuang:2012wp, Yan:1980uh}, as
\begin{widetext}
\begin{equation}
\begin{split}
    \frac{d\Gamma}{dm_{\pi\pi}d\cos{\theta_{\pi^+}}} =& \cos^{2}{\theta_{\rm mix}}
    \frac{dG_{\alpha}}{d m_{\pi\pi}d\cos{\theta_{\pi^+}}} 
    + \sin^{2}{\theta_{\rm mix}} \frac{dH_{\beta}}{d m_{\pi\pi}d\cos{\theta_{\pi^+}}}
\end{split}
\end{equation}
\end{widetext}
where $\theta_{\rm mix}$ is the mixing angle between $\psi(3686)$ and $\psi(3770)$. $\alpha$ and $\beta$ are phenomenological constants. 
%$G_{\alpha}$ and  $H_\beta$ are the corresponding S-wave and D-wave  decay widths, respectively.
$G_{\alpha}$ and $H_\beta$, representing the S-wave and D-wave decay widths respectively, are also constructed based on chiral symmetry to describe the hadronization amplitude for $\pi^+ \pi^-$.

In this case,  we observed a strong correlation between the mixing angle $\theta_{\rm mix}$ and the parameter $\beta$.
Fixing $\theta_{\rm mix}=12^{\circ}$ as determined from the ratio of electronic widths~\cite{Kuang:2006me}, the fit (Fig.~\ref{fig:totalfit}) reproduces the threshold enhancement but overshoots the spectrum near 0.55 GeV/$c^2$, resulting in a $\chi^2$/ndf of 308.5 without FSI, which improves to 9.2 when FSI is included.
However, the $\chi^2$/ndf in the $M(\pi^+\pi^-)\in[0.279, 0.325]~{\rm GeV}/c^2$ range is 33.7, which is significantly larger than the value of 1.5 under the resonance hypothesis as shown in Fig.~\ref{fig:pipifit}.

To account for the FSI contribution, the same FSI amplitude as described in Eq.~\ref{eq2} was applied to $G_\alpha$. The resulting fit yields 
$\alpha = -0.0884 \pm 0.0001_{\rm sta.} \pm 0.0016_{\rm sys.}$, 
$\beta = 0.29900 \pm 0.00620_{\rm sta.} \pm 0.02542_{\rm sys.}$, 
$\gamma = -3.6357 \pm 0.0055_{\rm sta.} \pm 0.1891_{\rm sys.}$.
As shown in Fig.~\ref{fig:totalfit}, the fitted projection is in agreement with the data, which implies that QCDME combined with FSI provides a significantly improved description of the data. 

Sources of systematic uncertainties on the fitting parameters are summarized in Table~\ref{tab:casec}. The uncertainties come from data-MC differences (MDC tracking, kinematic fit, and resolution), as well as from the requirements on $\cos{\theta}$ and the momentum of leptons ($P_l$). The data and MC efficiency difference for MDC tracking is studied using control samples with $J/\psi\rightarrow\pi^+\pi^-\pi^0$ and $J/\psi\rightarrow p \bar{p} \pi^+\pi^-$. The difference in fit results between the efficiencies with and without correction is taken as the systematic uncertainty. In accordance with the previous BESIII publication~\cite{BESIII:2012mpj}, the efficiency with the track helix correction is taken as the nominal value, and the difference in the fit results between the efficiencies with and without this correction is assigned as the systematic uncertainty from the kinematic fit. 
For the systematic uncertainty from the 1C kinematic fit (constraining the lepton-pair mass to the nominal $J/\psi$ mass), candidate events are selected directly from a 4C fit (energy-momentum conservation only), and the observed variations are assigned as systematic uncertainties.
For the requirements on $\cos{\theta}$ of the leptons, the discrepancies in the fit parameter, with and without this selection criterion, are regarded as the systematic uncertainties.
For the momentum requirement of the leptons, the threshold is varied from $0.15~{\rm GeV}/c$ to $0.19~{\rm GeV}/c$ in steps of $0.01~{\rm GeV}/c$, and the maximum change in the results is regarded as the systematic uncertainty. For the unfolding method, the number of iterations is varied from two to five, and the maximum change in the results is taken as the systematic uncertainty. 
For the resolution, the MC resolution is increased by 10\%, and the changes in the fit results are regarded as systematic uncertainties. 
The uncertainty due to the fit range is evaluated by varying the fit ranges.
The systematic uncertainty from the binning strategy is found to be negligible.

\begin{table}[htp]
    \caption{Summary of systematic sources and their contributions to the fitted parameters (in percent).}
    \centering
    \setlength{\tabcolsep}{2.4mm}{
    \begin{tabular}{c|c c c c c}
    \hline
        Sources& $\alpha$ & $\beta$ & $\gamma$ & $m_X$ & $\Gamma_X$ \\
    \hline
        MDC tracking & 0.2 & 0.9 & 0.1 & 0.4 & 1.8\\
        Kinematic fit & 1.6 & 7.5 & 4.8 & - & -\\
        $J/\psi$ 1C kinematic fit & 0.7 & 1.1 & 1.3 & - & -\\
        $\cos{\theta}$ requirement & 0.3 & 0.1 & 0.4 & - & -\\
        $P_{l}$ requirement & 0.1 & 1.3 & 1.2 & - & -\\
        Unfolding & 0.1 & 1.4 & 0.8 & - & - \\
        Resolution & 0.4 & 3.2 & 0.3 & 0.4 & 1.2\\
        Fit range & - & - & - & 0.7 & 0.9\\
    \hline
        Total & 1.8 & 8.5 & 5.2 & 0.9 & 2.3 \\
    \hline
    \end{tabular}}
    \label{tab:casec}
\end{table}

In summary, based on an unprecedented sample of $3.7\times 10^7$ $\psi(3686)\rightarrow\pi^{+}\pi^{-}J/\psi$ candidates selected from $(2712.4\pm14.4)\times 10^{6}$ $\psi(3686)$ decays collected with the BESIII detector, 
%a resonance-like structure
an enhancement near threshold in the $\pi^+\pi^-$ mass spectrum is clearly observed. A fit with a Breit-Wigner function yields a mass of $282.6\pm 0.4_{\rm sta.}\pm2.5_{\rm sys.}~{\rm MeV}/c^2$ and a width of $17.3\pm 0.8_{\rm sta.}\pm0.4_{\rm sys.}~{\rm MeV}$ for this anomalous structure. The measured width corresponds to a much shorter lifetime than that of a pionium state observed by DIRAC~\cite{DIRAC:2005hsg}. 
This discrepancy suggests that the observed structure originates from a different underlying mechanism, potentially involving strong interactions or exotic dynamics near the $\pi^+\pi^-$ threshold.
 
Further investigation of the $\pi^+\pi^-$ mass spectrum is performed using two distinct theoretical models. ChPT supplemented with FSI describes the $\pi^+\pi^-$ spectrum above 0.3 GeV/$c^2$ but fails near threshold.

QCDME with a mixing angle between S- and D-wave of $12^\circ$~\cite{Kuang:2006me} and FSI reproduces  the threshold enhancement and the high-mass region, suggesting that  $\psi(3686)$  is an  an admixture of S- and D-wave charmonium states. A noticeable discrepancy remains in the low-mass region ($\chi^2$/ndf $\approx 33.7$ in this region), suggesting that further investigations are necessary to clarify the underlying physics in this regime. The Adler zero requirement, which predicts a vanishing decay amplitude for $\psi(3686)\to\pi\pi J/\psi$ process~\cite{Pham:1975zq} could further help to explain the dip observed around 0.3 GeV/$c^2$.
Clarifying this feature will deepen our understanding of low-energy QCD and the role of chiral symmetry in hadronic transitions.

The BESIII Collaboration thanks the staff of BEPCII (https://cstr.cn/31109.02.BEPC) and the IHEP computing center for their strong support. This work is supported in part by National Key R\&D Program of China under Contracts Nos. 2025YFA1613900, 2023YFA1606000, 2023YFA1606704; National Natural Science Foundation of China (NSFC) under Contracts Nos. 11635010, 11935015, 11935016, 11935018, 12025502, 12035009, 12035013, 12061131003, 12192260, 12192261, 12192262, 12192263, 12192264, 12192265, 12221005, 12225509, 12235017, 12361141819; the Chinese Academy of Sciences (CAS) Large-Scale Scientific Facility Program; the Strategic Priority Research Program of Chinese Academy of Sciences under Contract No. XDA0480600; CAS under Contract No. YSBR-101; 100 Talents Program of CAS; The Institute of Nuclear and Particle Physics (INPAC) and Shanghai Key Laboratory for Particle Physics and Cosmology; ERC under Contract No. 758462; German Research Foundation DFG under Contract No. FOR5327; Istituto Nazionale di Fisica Nucleare, Italy; Knut and Alice Wallenberg Foundation under Contracts Nos. 2021.0174, 2021.0299; Ministry of Development of Turkey under Contract No. DPT2006K-120470; National Research Foundation of Korea under Contract No. NRF-2022R1A2C1092335; National Science and Technology fund of Mongolia; Polish National Science Centre under Contract No. 2024/53/B/ST2/00975; STFC (United Kingdom); Swedish Research Council under Contract No. 2019.04595; U. S. Department of Energy under Contract No. DE-FG02-05ER41374.

\bibliographystyle{apsrev4-2}
\bibliography{main}

\end{document}

%% file: authorlist_2025-07-02.tex
%% Saved at => 2025-07-02
M.~Ablikim$^{1}$\BESIIIorcid{0000-0002-3935-619X},
M.~N.~Achasov$^{4,b}$\BESIIIorcid{0000-0002-9400-8622},
P.~Adlarson$^{78}$\BESIIIorcid{0000-0001-6280-3851},
X.~C.~Ai$^{83}$\BESIIIorcid{0000-0003-3856-2415},
R.~Aliberti$^{37}$\BESIIIorcid{0000-0003-3500-4012},
A.~Amoroso$^{77A,77C}$\BESIIIorcid{0000-0002-3095-8610},
Q.~An$^{74,60,\dagger}$,
Y.~Bai$^{59}$\BESIIIorcid{0000-0001-6593-5665},
O.~Bakina$^{38}$\BESIIIorcid{0009-0005-0719-7461},
Y.~Ban$^{48,g}$\BESIIIorcid{0000-0002-1912-0374}
% ...
\\
\vspace{0.2cm}
(BESIII Collaboration)\\
\vspace{0.2cm} {\it
$^{1}$ Institute of High Energy Physics, Beijing 100049, People's Republic of China\\
% ...
$^{4}$ Budker Institute of Nuclear Physics SB RAS (BINP), Novosibirsk 630090, Russia\\
% ...
$^{37}$ Johannes Gutenberg University of Mainz, Johann-Joachim-Becher-Weg 45, D-55099 Mainz, Germany\\
$^{38}$ Joint Institute for Nuclear Research, 141980 Dubna, Moscow region, Russia\\
% ...
$^{48}$ Peking University, Beijing 100871, People's Republic of China\\
% ...
$^{59}$ Southeast University, Nanjing 211100, People's Republic of China\\
$^{60}$ State Key Laboratory of Particle Detection and Electronics, Beijing 100049, Hefei 230026, People's Republic of China\\
% ...
$^{74}$ University of Science and Technology of China, Hefei 230026, People's Republic of China\\
% ...
$^{77}$ University of Turin and INFN, (A)University of Turin, I-10125, Turin, Italy; (B)University of Eastern Piedmont, I-15121, Alessandria, Italy; (C)INFN, I-10125, Turin, Italy\\
$^{78}$ Uppsala University, Box 516, SE-75120 Uppsala, Sweden\\
% ...
$^{83}$ Zhengzhou University, Zhengzhou 450001, People's Republic of China\\

\vspace{0.2cm}
$^{\dagger}$ Deceased\\
% ...
$^{b}$ Also at the Novosibirsk State University, Novosibirsk, 630090, Russia\\
% ...
$^{g}$ Also at State Key Laboratory of Nuclear Physics and Technology, Peking University, Beijing 100871, People's Republic of China\\
% ...

}
%% ends here %%